\documentclass{article}
\usepackage{graphicx,subfigure}
\usepackage{amsmath,amsfonts}
\usepackage{enumerate}
\usepackage{hyperref}
\newtheorem{Remark}{Remark}
\newtheorem{Definition}{Definition}
\newtheorem{Theorem}{Theorem}

\DeclareMathOperator{\tr}{tr}

\title{Jacobi stability of circular orbits around conformally invariant  Weyl gravity black holes}

\author{Cristina Blaga\\Babe\c{s}-Bolyai University\\Faculty of Mathematics and Computer Science\\
	Kog\u{a}lniceanu Street 1, 600410, Cluj-Napoca, Romania\\
	Email: cristina.blaga@ubbcluj.ro\\
Paul A. Blaga\\Babe\c{s}-Bolyai University\\Faculty of Mathematics and Computer Science\\ Kog\u{a}lniceanu Street 1, 600410, Cluj-Napoca, Romania\\
Email: aurel.blaga@ubbcluj.ro}

\begin{document}
	\maketitle
	
\begin{abstract}
Weyl conformal gravity was originally proposed in the early twentieth century as an attempt to unify gravitation and electromagnetism. 
Since 1989, renewed interest in this fourth-order theory of gravity has emerged following the discovery of several exact black hole solutions. 
In this work, we investigate the timelike circular geodesics of a spherically symmetric Weyl black hole. The effective potential, the circular geodesics and their Jacobi and Lyapunov stability are discussed.
Our analysis provides new insights into the stability properties of Weyl black holes and the role of the free parameters appearing in their solutions.
\end{abstract}


\section{Introduction}

The general theory of relativity, proposed by Einstein in 1915,  was the first theoretical approach that succeeded in describing gravity in terms of the geometry of the surrounding spacetime. The geometric foundations of general relativity are represented by Riemannian geometry, which provides a powerful mathematical tool that allows for a deep understanding of gravitational interactions. A very important step in the evolution of gravitational theories was the derivation of the gravitational field equations from a variational principle, which was done by Hilbert. The theory of general relativity gives an extremely good description of gravitational phenomena in the Solar System, such as the perihelion precession of the planets, the bending of light by the Sun, or the Shapiro time-delay effect~\cite{Will,Mar}. A fundamental prediction of general relativity, the existence of gravitational waves, was also confirmed observationally ~\cite{grad1,grad2}.

Despite its successes, general relativity (GR) faces challenges at astrophysical and cosmological scales. Type Ia supernovae observations indicate an accelerated expansion of the Universe~\cite{acc1,acc2}, usually attributed to dark energy or a cosmological constant~\cite{Amen}, whose nature remains unknown. At galactic scales, rotation curves require dark matter~\cite{DM1,DM2}, which has not been directly detected. In addition, the $\lambda$ Cold Dark Matter model is affected by the Hubble tension, i.e., the discrepancy between Cosmic Microwave Background observations and low-redshift determinations of $H_0$~\cite{HTens}. These issues motivate extensions or modifications of general relativity.

Three years after the gravitational field equations were obtained by Einstein and Hilbert, Weyl~\cite{W18a,W18b} proposed an interesting and important generalization of the Riemannian geometry. In his approach, Weyl was guided by the idea of formulating a unified theory of the electromagnetic field and gravity. The Maxwell equations are conformally invariant in vacuum. Weyl proposed that the gravitational field equations must have the same symmetry, and thus all the laws of physics must be conformally invariant. Weyl introduced the principle of conformal invariance in a general way through the construction of a new geometry. The geometry proposed by Weyl is nonmetric, with the covariant derivative of the metric tensor not vanishing identically, a situation that makes Weyl geometry very different from the Riemannian one.  Moreover, the length of vectors changes during their parallel transport. Einstein strongly criticised the physical interpretation of Weyl’s theory, and its realistic (from the point of view of physical applications) nature. A discussion of the historical development of Weyl’s geometry, and of its recent physical applications, can be found in~\cite{Scholtz}. 

Weyl's idea of conformal invariance has attracted a lot of interest. An action based on a conformally invariant gravitational action, formulated with the help of the conformally invariant Weyl tensor $C_{\alpha \beta \gamma \delta}$ was proposed by Rudolf Bach~\cite{B21}.  A conformally invariant gravitational theory, defined in Riemannian geometry, with action given by $I_W =-\alpha\int d^4x\sqrt{-g}
C_{\mu\nu\rho\sigma}C^{\mu\nu\rho\sigma}$,
where $C_{\mu\nu\rho\sigma}$ is the Weyl tensor and $\alpha$ is a dimensionless constant, was introduced in~\cite{MK89},  and further investigated in~\cite{KM91,M92,M96}. The gravitational theories having actions obtained by using the Weyl tensor are called conformally invariant, or Weyl-type gravity theories. An exact vacuum static spherically symmetric solution of the conformally invariant Weyl gravity theory, with metric of the form $ds^2=B(r)dt^2-B^{-1}(r)dr^2-r^2\left(d\theta ^2+\sin ^2 \theta d\phi^2\right)$, where  
$
B(r)=1-3\beta \gamma -\frac{\beta\left(2-3\beta \gamma\right)}{r}+\gamma r+kr^2,
$
with $\beta$, $\gamma$ and $k$ being constants, obtained by Mannheim and Kazanas in 1989~\cite{MK89}. This solution, as well as conformal Weyl gravity, could represent a possible solution to the dark matter problem, because it can explain the observational data without the need to introduce dark matter in the form of a physical matter component~\cite{MK89}. 

The idea that scale invariance could play a fundamental role in gravitational physics has a long history. Early work by Dirac
~\cite{Dirac73} suggested that long-range interactions and gravitational coupling might be associated with a scalar field and the spontaneous breaking of scale symmetry, implying that dimensionful constants could arise dynamically rather than being fundamental. Such ideas naturally lead to conformal (Weyl-invariant) extensions of gravity, in which the action is constructed from the square of the Weyl tensor and is invariant under local scale transformations. A distinctive feature of these theories is the absence of any intrinsic mass scale at the classical level, so that physically relevant scales must emerge dynamically.

In recent years, significant progress has been made in understanding how local conformal symmetry can be embedded in realistic physical frameworks. In particular, Ghilencea and Lee~\cite{Ghil19} showed that Weyl gauge symmetry can be consistently implemented in the Standard Model and spontaneously broken, leading to the dynamical generation of the Planck scale and other mass parameters. This provides a concrete realization of the idea that scale invariance may underlie fundamental physics while still allowing for the emergence of observable scales.

At the same time, black hole solutions in conformal gravity have been the subject of increasing interest. In the context of fourth-order Weyl gravity, the Mannheim–Kazanas metric, which will be discussed in this paper, represents a prototypical static and spherically symmetric vacuum solution. More recently, this line of research has been extended to higher-derivative conformal gravity models, where additional curvature invariants can lead to a richer solution space and modified black hole properties. For example, Lessa et al.~\cite{lessa25} investigated black hole configurations in extended Weyl-invariant theories, highlighting how higher-order terms can affect horizon structure and spacetime geometry.

Weyl black holes are studied because they provide a framework for testing conformal gravity and identifying observable deviations from standard general relativity~\cite{EP98,SKS12}. The Mannheim--Kazanas metric contains additional integration constants compared to GR, leading to a broader class of spacetimes arising from its parameter space. The null geodesics of these spacetimes were investigated in~\cite{TH20}. In this paper, we study the timelike geodesics of a static, spherically symmetric Weyl black hole.  

The {\em KCC theory}, introduced by Kosambi~\cite{Ko33}, Cartan~\cite{Ca33}, and Chern~\cite{Ch39}, investigates the equivalence of certain classes of second-order ODE systems on manifolds and introduces a set of tensor fields, used to characterize the equivalence. {\em {Jacobi stability}} generalizes the concept of stability of the geodesic flow on a differentiable manifold. It provides a measure of the robustness and adaptability of a dynamical system to variations in its internal parameters and environmental conditions. It turns out that the Jacobi stability is closely related to the KCC theory, which provides a criterion for the Jacobi stability. 

The relation between Jacobi stability and Lyapunov stability of a dynamical system was studied by Boehmer et al.~\cite{rev}. Their conclusion was that, in general, Lyapunov and Jacobi stability are not equivalent. Therefore, Abolghasem~\cite{X2,X3} considered the stability of circular orbits in a central force field and in a Schwarzschild spacetime, looking for \emph{robust arrest regions}, regions where the trajectories are both Lyapunov and Jacobi stable. He demonstrated that for circular orbits in a central force field or in a Schwarzschild spacetime, the two types of stability are equivalent. 
Later, Jacobi and Lyapunov stability of circular timelike geodesics around a spherically symmetric dilaton black hole was analysed in~\cite{BBH2}. For other investigations of the application of the KCC theory for the study of the stability of various mathematical and physical systems, see~\cite{Ha6, BBH1}, and references therein.

It is the goal of the present work to investigate the Jacobi and linear stability of geodesics in the exact black hole geometry of the conformally invariant Weyl gravity, obtained in~\cite{MK89}. As a first step in our study, we obtain the geodesic equations of motion and the expression of the effective potential for a massive particle moving in the gravitational field of the Weyl gravity black hole. Then the stability of the trajectories of a massive particle moving in this potential is analyzed by using both the linear Lyapunov and the Jacobi stability approach. As a main result of our analysis, we find that for the conformally invariant Weyl gravity black hole solution, both stability methods predict the same stability~behavior.

The present paper is organized as follows.  We review the basics of the Jacobi and Lyapunov stability theories in Section~\ref{sect1}. We introduce the exact Weyl gravity black hole solution in Section~\ref{sect2}, where the properties of the horizon of the black hole are discussed. The geodesic equations, the effective potential, and the circular orbits of a test particle in motion around the black hole are obtained in Section~\ref{sect3}. The dynamical system analysis of the geodesics is performed in Section~\ref{sect4}. We briefly discuss our results and conclude our work in Section~\ref{sect5}.

\section{Jacobi Stability and Lyapunov Stability}\label{sect1}

In the present section, we briefly review the foundations of the KCC theory, Jacobi and Lyapunov stability, with emphasis on the relation between the two stability theories. 

\subsection{Geometric Framework of KCC Theory}

In this section, we summarize the principal concepts and results from KCC theory that will be employed in the subsequent analysis. For further details, see~\cite{An03}.

Let $\mathcal{M}$ be a real, smooth, $n$-dimensional manifold, and let $T\mathcal{M}$ denote its tangent bundle. 
We consider local coordinates: 
\begin{equation}
	(x^{i}) = (x^{1}, x^{2}, \ldots, x^{n}), \qquad (y^{i}) = (y^{1}, y^{2}, \ldots, y^{n}),
\end{equation}
where:
\begin{equation}
	(y^{i}) = \left( \frac{dx^{1}}{dt}, \frac{dx^{2}}{dt}, \ldots, \frac{dx^{n}}{dt} \right),
\end{equation}
and $t$ represents the time.  
Thus, $(x^{i}, y^{i}, t)$ form a system of $2n + 1$ coordinates on an open, connected subset $\Omega$ of the Euclidean space $\mathbb{R}^{n} \times \mathbb{R}^{n} \times \mathbb{R}$.

Consider the system of second-order differential equations:
\begin{equation}
	\frac{d^{2}x^{i}}{dt^{2}} + 2 G^{i}(x^{j}, y^{j}, t) = 0, 
	\qquad i = \overline{1, n}, 
	\label{EM}
\end{equation}
where $G^{i}(x^{j}, y^{j}, t)$ are $C^{\infty}$ functions defined in a neighborhood of the initial conditions $(x_{0}, y_{0}, t_{0}) \in \Omega$.

In 1933, D. Kosambi~\cite{Ko33} gave an answer to the following problem: Find the coordinate transformations on $M$ that leave invariant the solution curves of the system~(\ref{EM}). It turned out that this amounted to the invariance of a set of tensor fields. Cartan~\cite{Ca33}, acting as referee for Kosambi's paper, noticed that the paper was incomplete and wrote a short paper that completely solved the problem, which was published in the same issue of Mathematische Zeitschrift. Finally, Chern, in 1939,~\cite{Ch39} reformulated the theory in the language of the equivalence of differential forms, created, as well, by Cartan.

One of the main ideas behind the KCC theory is that the Equation~(\ref{EM}) can be written as the geodesic equations for a well-chosen connection and then we can adapt tools from Riemannian geometry to deal with the trajectories of these equations.

One of the classical results of Riemannian geometry is that the separation between neighbouring geodesics is described by \emph{{Jacobi vector fields}}, which verify the Jacobi equation. By studying the geodesics of surfaces, Jacobi discovered that for surfaces with positive Gauss curvature, the geodesics refocus, i.e., two initially close geodesics remain close, or in other words, the geodesics are \emph{{stable}} with respect to small perturbations. The results of Jacobi were extended to arbitrary Riemannian manifolds by T. Levi-Civit\`a (see~\cite{LC27}), who introduced the notion of \emph{{geodesic deviation}}. In parallel, J. Synge \cite{Syn27} developed a geometrical theory of dynamics, using, among other stability notions, the stability we shall define below, the \emph{{Jacobi stability}} for Riemannian geodesics.

The term \emph{Jacobi stability} for general systems of the form~(\ref{EM}) was introduced by Peter Antonelli and his collaborators in the late eighties and early nineties (see~\cite{AnIM93}). They noticed that the Jacobi stability is connected to one of the KCC invariant tensors associated with the system of equations, namely the so-called \emph{deviation curvature tensor}.

We are not going to enter into any geometrical details here; we shall simply review what we need for handling our problem. Details about the KCC theory can be found in~\cite{An03}, while for Jacobi stability, in~\cite{rev,An01}, and references therein.

Exactly as a first order ODEs system on a manifold $\mathcal{M}$ is described by a vector field on $\mathcal{M} $, to a second order ODEs system we associate a vector field on the tangent bundle, $T\mathcal{M}$, called a \emph{semispray}. In the case of the system
~\eqref{EM}, the semispray is given by:
\begin{equation}
	S = y^{i} \frac{\partial}{\partial x^{i}} 
	- 2 G^{i}(x^{j}, y^{j}, t) \frac{\partial}{\partial y^{i}},
\end{equation}
which  induces a nonlinear connection $N^{i}_{j}$ defined by:
\begin{equation}
	N^{i}_{j} = \frac{\partial G^{i}}{\partial y^{j}}.
\end{equation}

Let $x^{i}(t)$ be a trajectory of the system~\eqref{EM}, and consider a nearby trajectory:
\begin{equation}
	\tilde{x}^{i}(t) = x^{i}(t) + \eta\, \xi^{i}(t), 
	\label{var}
\end{equation}
where $|\eta|$ is a small parameter and $\xi^{i}(t)$ are the components of a contravariant vector field defined along $x^{i}(t)$. 

Substituting~\eqref{var} into~\eqref{EM} and taking the limit $\eta \to 0$ yields:
\begin{equation}
	\frac{d^{2}\xi^{i}}{dt^{2}} + 2 N^{i}_{j} \frac{d\xi^{j}}{dt} 
	+ 2 \frac{\partial G^{i}}{\partial x^{j}} \xi^{j} = 0.
	\label{def}
\end{equation}

Equation~\eqref{def} can be expressed in a covariant form as follows:
\begin{equation}
	\frac{D^{2}\xi^{i}}{dt^{2}} = P^{i}_{j} \xi^{j},
	\label{JE}
\end{equation}
where the operator $\dfrac{D}{dt}$ denotes the KCC-covariant differential,
defined by:
\begin{equation}
	\frac{D\xi_i}{dt}=\frac{d\xi_i}{dt}+N_j\xi^j
\end{equation}
and
\begin{equation}\label{deviat}
	P^{i}_{j} 
	= -2 \frac{\partial G^{i}}{\partial x^{j}}
	- 2 G^{l} G^{i}_{jl}
	+ y^{l} \frac{\partial N^{i}_{j}}{\partial x^{l}}
	+ N^{i}_{l} N^{l}_{j}
	+ \frac{\partial N^{i}_{j}}{\partial t},
\end{equation}
with $G^{i}_{jl} \equiv \tfrac{\partial N^{i}_{j}}{\partial y^{l}}$ representing the \emph{{Berwald connection}}.

Equation~\eqref{JE} is known as the \emph{Jacobi equation}, and the tensor $P^{i}_{j}$ is referred to as the \emph{{second KCC invariant}} or the \emph{deviation curvature tensor}.

\begin{Definition}
	If the system of ordinary differential Equation~\eqref{EM}
	satisfies the initial conditions: 
	\begin{equation}
		\left| \left| x^{i}\left( t_{0}\right) -
		\tilde{x}^{i}\left( t_{0}\right) \right| \right| =0\,, \,\, \left| \left| \dot{x}
		^{i}\left( t_{0}\right) -\dot{\tilde{x}}^{i}\left( t_{0}\right) \right| \right|
		\neq 0,
	\end{equation}	
	with respect to the norm $\left| \left| .\right| \right| $, induced
	by a positive definite inner product, then the trajectories of the system~\eqref{EM}
	are called \emph{Jacobi stable} if and only if the real parts of the eigenvalues of the
	deviation tensor $P_{j}^{i}$ are strictly negative everywhere, and \emph{Jacobi
			unstable}, if this condition does not hold.  
\end{Definition}

\subsection{Lyapunov Stability}

In this section, we summarize the main concepts and results from Lyapunov stability theory that we will use in the following. For more details, see~\cite{H67}.
\begin{Definition}
	A \emph{{fixed point}} of a system of autonomous ordinary differential equations (ODEs):
	\begin{equation}
		\dot{x} = f(x),
		\label{sed}
	\end{equation}
	is a point $\bar{x} \in \mathbb{R}^n$ such that $f(\bar{x}) = 0$, where $f$ is a $C^{1}$ vector field on $\mathbb{R}^{n}$.
\end{Definition}

\begin{Remark}
	Fixed points are also referred to as \emph{{equilibrium points}}.
\end{Remark}

\begin{Definition}
	Let $\bar{x}$ be a fixed point of the differential Equation~\eqref{sed}. The point $\bar{x}$ is called a \emph{{hyperbolic fixed point}} if 
	$\operatorname{Re}(\lambda_i) \neq 0$ for all eigenvalues $\lambda_i$ of the Jacobian matrix of $f(x)$ evaluated at $\bar{x}$.  
	Otherwise, $\bar{x}$ is said to be \emph{{non-hyperbolic}}.
\end{Definition}

\begin{Definition}
	The fixed point $\bar{x}$ is said to be \emph{stable} (or \emph{{Lyapunov stable}}) if, for every $\varepsilon > 0$, there exists $\delta = \delta(\varepsilon) > 0$ such that for any solution $y(t)$ of~\eqref{sed} satisfying:
	\begin{equation}
		\| \bar{x}(t_0) - y(t_0) \| < \delta, \quad t_0 \in \mathbb{R},
	\end{equation}
	we have:
	\begin{equation}
		\| \bar{x}(t) - y(t) \| < \varepsilon, \quad \forall\, t > t_0.
	\end{equation}
	
	A solution that is not stable is said to be \emph{unstable}.
\end{Definition}

\begin{Theorem}[Lyapunov Stability Theorem]
	Consider the vector field $\dot{x} = f(x)$, with $x \in \mathbb{R}^{n}$. Let $\bar{x}$ be an equilibrium point of the system, and assume there exists a  $C^{1}$  function
	$V : U \to \mathbb{R}$ defined on some neighborhood $U$ of $\bar{x}$ such that:
	
	\begin{enumerate}
		\item $V(\bar{x}) = 0$ and $V(x) > 0$ for all $x \neq \bar{x}$;
		\item $\dot{V}(x) \le 0$ for all $x \in U \setminus \{\bar{x}\}$.
	\end{enumerate}
	Then the equilibrium point $\bar{x}$ is stable.
\end{Theorem}

\begin{Definition}
	The function $V(x)$ from the above theorem is called a \emph{Lyapunov function}.
\end{Definition}

A qualitative analysis of a dynamical system begins by locating its fixed points.  
If $f(x)$ is of class $C^{1}$, the local behavior of the orbits near a fixed point $\bar{x}$ can be approximated by linearizing the vector field:
\begin{equation}
	f(x) \approx Df(\bar{x})(x - \bar{x}),
	\label{lin}
\end{equation}
where $Df(\bar{x})$ denotes the Jacobian matrix of $f$ evaluated at $\bar{x}$.

\begin{Definition}
	The system~\eqref{lin} is called the \emph{{linearization}} of the differential equation at $\bar{x}$.
\end{Definition}

\begin{Theorem}[Hartman--Grobman Theorem]
	Consider the differential equation $\dot{x} = f(x)$, with $x \in \mathbb{R}^{n}$ and $f \in C^{1}$.  
	If $\bar{x}$ is a \emph{{hyperbolic}} fixed point of the system, then there exists a neighborhood of $\bar{x}$ on which the flow of the nonlinear system is \emph{{topologically equivalent}} to the flow of its linearization at $\bar{x}$.
\end{Theorem}

\begin{Remark}
	The classification of fixed points can be carried out by studying the eigenvalues of the Jacobian matrix of the linearized vector field at those points.
\end{Remark}

\subsection{Jacobi and Lyapunov Stability}

The relationship between Lyapunov and Jacobi stability for two-dimensional systems was discussed in detail by Boehmer {et al.}~\cite{rev}.  
They established a correspondence between the two notions of stability: the \emph{{linear stability}}, which depends on the signs of the eigenvalues of the Jacobian matrix $J$, and the \emph{{Jacobi stability}}, which is determined by the signs of the eigenvalues of the deviation curvature tensor $P^{i}_{j}$ evaluated at the same point.

Consider the autonomous system of ODEs:
\begin{equation}\label{ode}
	\begin{cases}
		\dot{u} = f(u, v),\\
		\dot{v} = g(u, v),
	\end{cases}
\end{equation}
with an equilibrium point at $(0, 0)$, i.e., $f(0,0) = g(0,0) = 0$.  
The Jacobian matrix of~\eqref{ode} is given by:
\begin{equation}
	J(u, v) =
	\begin{pmatrix}
		f_{u} & f_{v} \\
		g_{u} & g_{v}
	\end{pmatrix},
\end{equation}
where subscripts denote partial derivatives with respect to the indicated variables.

The characteristic equation is:
\begin{equation}
	\lambda^{2} - (\operatorname{tr} A)\lambda + \det A = 0,
\end{equation}
where $\operatorname{tr} A$ and $\det A$ denote the trace and determinant of the matrix $A := J|_{(0,0)}$, \mbox{respectively}.

The linear stability of the equilibrium $(0,0)$ depends on the signs of the discriminant:
\begin{equation}
	\Delta =\left(\operatorname{tr}A\right)^2-4\det A.
\end{equation}

\begin{Theorem}
	Consider the ODE system~\eqref{ode} with a fixed point $M(0, 0)$ such that $g_{u}|_{(0,0)} \neq 0$.  
	Then the Jacobian $J$ evaluated at $M$ has complex eigenvalues if and only if $M$ is a \emph{{Jacobi stable~point}}.
\end{Theorem}

\begin{Remark}
	The easiest way to understand the relation between the two notions of stability is to consider some examples. We start with a system of  two linear first order ODEs with constant~\mbox{coefficients}:
	\begin{equation}\label{remeq1}
		\begin{cases}
			\dot{x}=ax+by,\\
			\dot{y}=cx+dy.
		\end{cases}	
	\end{equation}	
	
	We denote by:
	\begin{equation}\label{remeq2}
		A=
		\begin{pmatrix}
			a&b\\
			c&d
		\end{pmatrix}
	\end{equation}
	the matrix of this system.
	
	As it is well-known from the theory of ODEs, the system~(\ref{remeq1}) is equivalent to the linear, second-order ODE with constant coefficients (see~\cite{rev} for the discussion of the nonlinear case):
	\begin{equation}\label{remeq3}
		\ddot{x}-\left(\tr A\right) \dot{x}+\left(\det A\right) x=0.
	\end{equation}	
	
	{Now,}
	clearly, this is an equation of the same kind as~(\ref{EM}), with $n=1$, with:
	\begin{equation}\label{remeq4}
		G^1(x,\dot{x})\equiv G(x,\dot{x})=\frac{1}{2}\left(-\left(\tr A\right) \dot{x}+\left(\det A\right) x\right).
	\end{equation}
	
	{The}
	associated nonlinear connection has a single coefficient:
	\begin{equation}\label{remeq5}
		N_1^1(x,\dot{x})\equiv N(x,\dot{x})=\frac{\partial G}{\partial\dot{x} }=-\frac{1}{2}\tr A,
	\end{equation}
	while, as $N$ is constant, the only coefficient of the Berwald connection vanishes. As such; therefore, the only coefficient of the deviation tensor (see Equation~(\ref{deviat})) is:
	\begin{equation}
		P_1^1(x,\dot{x})\equiv P(x,\dot{x})=-2\frac{\partial G}{\partial x}+N^2, 
	\end{equation}
	or
	\begin{equation}\label{remeq6}
		P(x,\dot{x})=\frac{1}{4}\left(\tr A\right)^2-\det A=\frac{(a+d)^2}{4}-(ad-bc).
	\end{equation}
	
	{Let} 
	us denote, in the spirit of~\cite{HSD}, $p=(a+d)$ and $q=ad-bc$. 
	
	According to the definition we gave, the trajectories of the system are Jacobi stable iff $P<0$, or, which is the same:
	\begin{equation}
		p^2-4q<0.
	\end{equation}
	
	{On}
	the other hand, we can use $p$ and $q$ to characterize the origin, as an equilibrium point of the system~(\ref{remeq1}). We follow the books~\cite{H67,HSD}.
	
	First of all, we consider the characteristic equation of the matrix $A$, namely:
	\begin{equation}
		\lambda^2-p\lambda+q=0.
	\end{equation}
	
	{The} discriminant of this equation is: 
	\begin{equation}
		\delta=p^2-4q.
	\end{equation}
	
	{Thus,}
	the sign of $\delta$ is the sign of $P$, the deviation tensor. The eigenvalues of $A$ are:
	\begin{equation}
		\lambda_\pm =\frac{1}{2}\left(p\pm \sqrt{p^2-4q}\right).
	\end{equation}
	
	{Consider,} 
	first, the case $\delta <0$. The trajectory is then \emph{Jacobi stable}. Let us see what happens to the Lyapunov stability. The (common) real part of the eigenvalues is $p/2$. Thus, we have three~situations:
	
	\begin{enumerate}
		\item $p<0$, in this case, the solution is stable (we have a stable focus (\emph{{spiral sink}})).
		\item $p>0$, in this case, the solution is unstable (unstable focus or \emph{{spiral source}}).
		\item $p=0$, in this case the solution is periodic, we get a center.
	\end{enumerate}
	
	{If} 
	$\delta>0$, the system is Jacobi unstable. In what concerns the Lyapunov stability, we have, again, three situations:
	\begin{enumerate}
		\item If $q<0$, as $q$ is the product of the eigenvalues, it follows that the origin is a saddle point (since the eigenvalues have opposite signs).
		\item If $q>0$ and $p<0$, then both eigenvalues are negative, which means we have a stable node, so the system is Lyapunov stable.
		\item If $q>0$ and $p>0$, then both eigenvalues are positive, therefore the origin is an unstable node, i.e. the system is unstable.
	\end{enumerate}
	
	{Thus,} 
	we can easily construct some examples of equations with different kinds of behaviour in what concerns the stability. (The phase portraits for these equations are standard material in ODE theory and can be found, for instance, in~\cite{H67,HSD}, or in the paper~\cite{rev}.)
	
	\begin{enumerate}
		\item If we let $a=2, b=c=0$ and $d=1$, we have the $p=3, q=2$, hence $\delta =1>0$. Thus, the trajectories of the equation:
		\begin{equation}
			\ddot{x}-3\dot{x}+2x=0
		\end{equation}	
		are Jacobi unstable. On the other hand, the origin is an unstable node; therefore, we have  Lyapunov instability, as well.
		\item If $a=1,b=-1, c=1, d=1$, we have $p=2, q=2$, therefore $\delta=-6<0$, so the trajectories of the equation:
		\begin{equation}
			\ddot{x}-2\dot{x}+2x=0
		\end{equation}	
		are Jacobi stable, but the origin is an unstable focus, i.e., we have Lyapunov instability.
		\item If we take $a=-1, b=-1, c=1, d=-1$, then $p=-2, q=2$, while  $\delta=-6<0$. The second-order equation is:
		\begin{equation}
			\ddot{x}+2\dot{x}+2x=0.
		\end{equation}
		{This} 
		time the origin is a stable focus, and the trajectories are Jacobi stable, as $\delta<0$, so both types of stability are present.
		\item If we let $a=-1, b=c=0, d=-2$, then $p=-3, q=2$, hence the equation is:
		\begin{equation}
			\ddot{x}+3\dot{x}+2x=0.
		\end{equation}
		The trajectory is Jacobi unstable, because $\delta=1>0$, but the origin is a \emph{stable} node.
		\item As a final example, we can let $a=0, b=-1, c=1, d=0$. Thus, $p=0$, $q=1$. The equation is the equation of the harmonic oscillator:
		\begin{equation}
			\ddot{x}+x=0.
		\end{equation}
		The origin is, this time, a center, but the trajectories are Jacobi stable, as $\delta=-4<0$.
	\end{enumerate}
	
	{In}
	the last example, the trajectories are circles with the center at the origin. This example shows that, in the case of a Jacobi stable system, trajectories do not necessarily converge to a single point (or~originate from the same point).
\end{Remark}

\section{Black Hole Solutions in Conformally Invariant Weyl Gravity}\label{sect2}

The Weyl theory of gravity was introduced by Hermann Weyl in 1918 in an attempt to unify gravitation and electromagnetism~\cite{W18a, W18b}. 
Weyl’s idea was to supplement the spacetime metric with an additional four-dimensional vector field, the \emph{{Weyl vector}}, and to replace the Levi-Civit\`a connection with a new affine connection constructed from both the Levi-Civit\`a connection and the Weyl vector. 
All curvature tensors and their contractions (such as the Ricci tensor and the scalar curvature) are then defined with respect to this new connection. 
A distinctive feature of Weyl gravity is its \emph{{conformal invariance}}. As mentioned earlier, the theory of gravity we use is \emph{{not}} the original Weyl theory (or, rather, is a Weyl theory in which the supplementary vector field vanishes). We only keep from Weyl theory the conformal invariance, but all our curvature computations are done using the ordinary Levi-Civit\`a connection associated to the metric.

In 1921, Rudolf Bach proposed the following action for Weyl gravity~\cite{B21}:
\begin{equation}\label{act1}
    I_{W} = -\alpha \int d^{4}x \, \sqrt{-g} \, C_{\mu\nu\rho\sigma} C^{\mu\nu\rho\sigma},
\end{equation}
where $C_{\mu\nu\rho\sigma}$ is the \emph{{conformal Weyl tensor}}, defined as follows:
\begin{equation}
    C_{\mu\nu\rho\sigma} = R_{\mu\nu\rho\sigma}
    - \left( g_{\mu[\rho} R_{\sigma]\nu} - g_{\nu[\rho} R_{\sigma]\mu} \right)
    + \frac{1}{3} R g_{\mu[\rho} g_{\sigma]\nu}.
\end{equation}
$\alpha$ is a dimensionless parameter that acts as a coupling constant of the theory. (Unlike the Einstein-Hilbert action, where Newton’s constant introduces a mass scale, here no intrinsic scale appears in the fundamental action.
This reflects the scale (conformal) invariance of the theory.) See the review paper of Mannheim~\cite{M06} for more details about the physical significance of this constant.

Kazanas and Mannheim adopted this action for the conformal Weyl gravity, thought of as a quadratic curvature, conformally invariant gravity theory, constructed in the realm of Riemannian geometry.

It is easy to check that the Lagrangean can be written, in terms of curvature\linebreak quantities~as:
\begin{equation}
\mathcal{L}:=C_{\mu\nu\rho\sigma} C^{\mu\nu\rho\sigma}=R_{\mu\nu\rho\sigma} R^{\mu\nu\rho\sigma}-2 R_{\mu\nu}R^{\mu\nu}	+\frac{1}{3}R^2.
\end{equation}

{We} 
 are going to show that the term containing the ``square of the curvature tensor'', 
$R_{\mu\nu\rho\sigma} R^{\mu\nu\rho\sigma}$ can be safely removed from the Lagrangean, without changing the associated equations of motion.

We consider the so-called \emph{Gauss-Bonnet} term:
\begin{equation}
\mathcal{G}=R_{\mu\nu\rho\sigma} R^{\mu\nu\rho\sigma}-4 R_{\mu\nu}R^{\mu\nu}	+R^2.
\end{equation}

{We}
 have, then:
\begin{equation}
\mathcal{L}-\mathcal{G}=2R_{\mu\nu}R^{\mu\nu}-\frac{2}{3}R^2.
\end{equation}

{Thus,} 
 we can write:
\begin{equation}
\mathcal{L}=2\left(R_{\mu\nu}R^{\mu\nu}-\frac{1}{3}R^2\right)+\mathcal{G}.
\end{equation}

{In} 
 four dimensions,  the Gauss--Bonnet term is a total divergence; therefore, its variation is zero and we can drop it from the Lagrangian, getting the same equations of motion. In fact, it is known that the integral of the Gauss--Bonnet term is, up to a constant factor, nothing but the Euler--Poincar\'e characteristic of the spacetime~\cite{nak2003}, a \emph{{topological}} invariant of spacetime (an integer), therefore we can say that the Gauss--Bonnet term is a \emph{{topological term}}. In fact, the result that the Gauss--Bonnet term was a divergence was already known, in 1938, to Lanczos~\cite{Lan38}. See also the recent article of Condeescu and Micu~\cite{cond25} for a modern~approach.

Thus, we can use, as an action:
\begin{equation}
    I'_{W} = -2\alpha \int d^{4}x \, \sqrt{-g}
    \left( R_{\mu\nu} R^{\mu\nu} - \frac{1}{3} R^{2} \right).
\end{equation}

The metric for a static, spherically symmetric, and uncharged solution of the gravitational field equations was obtained by Mannheim and Kazanas~\cite{MK89}:
\begin{equation}\label{metr}
    ds^{2} = -B(r) \, dt^{2} + \frac{dr^{2}}{B(r)} 
    + r^{2} \left( d\theta^{2} + \sin^{2}\theta \, d\varphi^{2} \right),
\end{equation}
where:
\begin{equation}
    B(r) = 1 - \frac{\beta(2 - 3\beta\gamma)}{r} 
    - 3\beta\gamma + \gamma r - k r^{2},
\end{equation}
and $\beta$, $\gamma$, and $k$ are integration constants.
\begin{Remark}
	The fact that the family of spherically symmetric Mannheim--Kazanas black holes depends on three parameters is no accident. Unlike the Einstein field equations, the equations derived from the Bach action are of fourth order. In particular, the function $B$ is obtained as a solution of a fourth order ordinary differential equation. As such, we should have four integration constants. We have an extra degree of freedom from the conformal invariance of the solution. We used this extra degree of freedom when we singled out the form~(\ref{metr}) of the metric. 
	
	In the original paper~\cite{MK89}, Mannheim and Kazanas started with a general static, spherically symmetric line element:
\begin{equation}\label{ss}
ds^2=-b(\rho)dt^2+a(\rho)d\rho^2+\rho^2d\Omega^2.
\end{equation}	

{They} 
 notice that, by changing the radial coordinate $\rho$ after the rule:
\begin{equation}\label{ss1}
\rho=p(r)=-\dfrac{1}{{\displaystyle\int\dfrac{dr}{r^2[a(r)b(r)]^{1/2}}}}
\end{equation}
(\ref{ss}) becomes:
\begin{equation}\label{ss2}
ds^2=\frac{p^2(r)}{r^2}\left[-B(r)dr^2+\dfrac{1}{B(r)}dr^2+r^2d\Omega^2\right]\,.
\end{equation}

Now,
 as the field equations in Weyl gravity are conformally invariant, we can drop the factor $p^2(r)/r^2$ and we get a line element of the form~(\ref{metr}). The ordinary differential equation that the function $B$ has to satisfy (with this particular choice of the conformal scaling and radial coordinate) is of third order; therefore, we obtain a line element depending only on three arbitrary constants (see~\cite{MK89}).

Sometimes, a different conformal factor is used for specific purposes. For  instance, Turner and Horne~\cite{TH20} suggest that one can use a factor:
\begin{equation}
\Omega^2(r)=1+\frac{L^4}{r^4},
\end{equation}
where $L$ is a length scale parameter,
in front of the metric~(\ref{metr}) to eliminate the singularity at $r=0$.
\end{Remark}

Mannheim and Kazanas~\cite{MK89} pointed out that the parameter $\gamma$ measures the deviation of Weyl gravity from Einstein’s general relativity. 
The parameter $\gamma$ is extremely small and in the limit $\gamma \to 0$, the Schwarzschild solution in an (anti-)de~Sitter background is recovered. 
The parameter $\beta$ is associated with the mass of the source, while $k$ is related to the cosmological curvature scale.  
In~\cite{MK89}, the authors showed that for $\gamma \sim 10^{-28} \, \mathrm{cm}^{-1}$ and $\beta\gamma \ll 1$ the observed galactic rotation curves can be reproduced. 
In~\cite{M92}, Mannheim found that $k$ should be negative, with a representative value $k = -3.5 \times 10^{-60} \, \mathrm{cm}^{-2}$, as inferred from galactic rotation curve data~\cite{M96}.
In~\cite{MO13}, Mannheim and O'Brien derived a positive value for $k$ to fit the observed galactic rotation curves of 138 spiral galaxies. In~\cite{EP98} Edery and Paranjape obtained that a negative sign is required to increase the deflection of light on large distant scales. Features of spacetimes with Mannheim--Kazanas metric for different values of the parameters are analysed in~\cite{TH20}. 

Next, we consider negative values for $k$ and positive values for $\gamma$ and $\beta$, so that $\beta\gamma \ll 1$.   
In terms of dimensionless parameters $\tilde{r}=r/\beta$, $\tilde{\gamma}=\beta \gamma$ and $\tilde{k}=\beta^2 k$, the lapse function is:
\begin{equation}
    B(\tilde{r}) = 1 - \frac{2 - 3\tilde{\gamma}}{\tilde{r}} 
    - 3\tilde{\gamma} + \tilde{\gamma} \tilde{r} - \tilde{k} \tilde{r}^{2}.
\end{equation}

As $\tilde{r} \rightarrow 0^{+}$, the function $B(\tilde{r})$ tends to $\operatorname{sgn}(3\tilde{\gamma} - 2) \infty$, which equals $-\infty$, since $\tilde{\gamma} \ll 1$. As $\tilde{r}$ increases toward $+\infty$, the function $B(\tilde{r})$ diverges to $- \operatorname{sgn}(\tilde{k}) \infty$, which is equal to $+\infty$, since $\tilde{k} < 0$.   
The function $B(\tilde{r})$ is continuous on $(0, \infty)$; therefore, there exists $\tilde{r}_{0} \in (0, \infty)$ such that $B(\tilde{r}_{0}) = 0$. Horizons of the metric occur at coordinate singularities where $B(\tilde{r})=0$. A detailed discussion on horizons of the metric~(\ref{metr}) for different values of the parameters $\beta$, $\gamma$ and $k$ is done by Turner and Horne in~\cite{TH20}. 

As for the interval in which the horizon is, we see that the numerator of the lapse~\mbox{function}:
\begin{equation}
    P(\tilde{r}) = -\tilde{k} \tilde{r}^{3} + \tilde{\gamma} \tilde{r}^{2} + (1 - 3\tilde{\gamma})\tilde{r} + 3\tilde{\gamma} - 2, \label{P0}
\end{equation}
is strictly increasing for $\tilde{r}>0$. Indeed, its derivative: \begin{equation}
	    P'(\tilde{r}) = -3\tilde{k} \tilde{r}^{2} + 2\tilde{\gamma} \tilde{r} + 1 - 3\tilde{\gamma} \,, \label{P1}
	\end{equation}
	is positive $\forall \tilde{r}>0$, since $\tilde{k}<0$ and $\tilde{\gamma} \ll 1$. The unique root of $P(\tilde{r})=0$ satisfies $\tilde{r}_{0}<2$, because $P(2)=\tilde{\gamma}-8\tilde{k}>0$. This implies that the horizon $r_0$ is inside the Schwarzschild radius $r_{\text{S}}=2 \beta$ of the corresponding Schwarzschild black hole obtained for $\gamma=0$ and $k=0$. We note that this result generalizes the one obtained by Lungu and Dariescu in~\cite{LD23}, for $k=0$ and $\beta \gamma \ll 1$. 
	

	\section{Geodesics Equations, Effective Potential and Circular Orbits}\label{sect3}
	
	A free test particle moves around the black hole along a timelike geodesic. The geodesic equations can be derived from the Euler--Lagrange formalism. 
	
	The Lagrangian corresponding to the metric~\eqref{metr} is:
	\begin{equation}\label{lagr}
	    2\mathcal{L} = -B(r)\dot{t}^{2} + \frac{\dot{r}^{2}}{B(r)} + r^{2}\left(\dot{\theta}^{2} + \sin^{2}\theta \, \dot{\varphi}^{2}\right),
	\end{equation}
	where the overdot denotes differentiation with respect to the affine parameter $\tau$ along the geodesic. 
	The normalization of the Lagrangian is chosen such that $2\mathcal{L} = -1$ for timelike geodesics, $2\mathcal{L} = 0$ for null geodesics and $2\mathcal{L} = 1$ for spacelike geodesics.  
	
	\subsection{Euler--Lagrange Equations}
	
	The coordinates $t$ and $\varphi$ are cyclic, leading to two conserved quantities.  
	The first integral of motion is:
	\begin{equation}\label{ien}
	    B(r)\dot{t} = \text{constant} = E,
	\end{equation}
	which represents the conserved energy per unit mass of the test particle.  
	
	The second integral of motion is:
	\begin{equation}\label{imc}
	    r^{2}\sin^{2}\theta\,\dot{\varphi} = \text{constant},
	\end{equation}
	corresponding to the conservation of angular momentum.
	
	From the Euler--Lagrange equation for $\theta$, one obtains:
	\begin{equation}\label{ecteta}
	    \frac{d}{d\tau}\!\left(r^{2}\dot{\theta}\right) = r^{2}\sin\theta\cos\theta\,\dot{\varphi}^{2}.
	\end{equation}
	
	{If} 
	 $\theta = \pi/2$ and $\dot{\theta} = 0$ at some point, then $\ddot{\theta} = 0$ and thus $\theta = \pi/2$ along the entire geodesic.  
	Hence, the geodesics are planar, as in the Schwarzschild spacetime or in the Newtonian gravitational field.
	
	Restricting the motion to the equatorial plane ($\theta = \pi/2$), Equation~\eqref{imc} reduces to:
	\begin{equation}
	    r^{2}\dot{\varphi} = L,
	\end{equation}
	where $L$ is the conserved angular momentum per unit mass.
	
	The Euler--Lagrange equation for $r$ can be replaced by the constancy of the Lagrangian:
	\begin{equation}\label{et}
	    \left(\frac{dr}{d\tau}\right)^{2} + B(r)\!\left(\frac{L^{2}}{r^{2}} - \epsilon\right) = E^{2},
	\end{equation}
	where $\epsilon = -1, 0, +1$ correspond to timelike, null, and spacelike geodesics, respectively.
	
	\subsection{Effective Potential for Spherically Symmetric Weyl Black Holes}
	
	Comparing Equation~\eqref{et} with the Newtonian energy conservation law, one may interpret the second term on the left-hand side as an \emph{{effective potential}}, defined by:
	\begin{equation}\label{Veff}
	    V_{\text{eff}}(r) = \frac{1}{2}B(r)\left(\frac{L^{2}}{r^{2}} - \epsilon\right).
	\end{equation}
	
	{For} 
	 timelike geodesics ($\epsilon = -1$), this becomes:
	\begin{equation}\label{Ve}
	    V_{\text{eff}}(r) = 
	    \frac{\left[-k r^{3} + \gamma r^{2} + (-3\beta\gamma + 1)r + 3\beta^{2}\gamma - 2\beta\right](L^{2} + r^{2})}{2r^{3}}\,\,.
	\end{equation}
	
	For positive $r$, the equation $V_{\text{eff}}(r)=0$ has a unique solution, the positive root $r_0$ of the lapse function $B(r)$, because the factor $L^{2} + r^{2}$ is strictly positive for all $r>0$.
	
	By introducing the dimensionless parameter $\tilde{L}=L/\beta$, the effective potential~(\ref{Ve}) becomes:
	\begin{equation}\label{Vef}
	    V_{\text{eff}}(\tilde{r}) = 
	    \frac{\left[-\tilde{k} \tilde{r}^{3} + \tilde{\gamma} \tilde{r}^{2} + (1-3\tilde{\gamma})\tilde{r} + 3\tilde{\gamma} - 2\right](\tilde{L}^{2} + \tilde{r}^{2})}{2\tilde{r}^{3}}\,\,.
	\end{equation}
	
	{As}
	 $\tilde{r} \rightarrow 0^{+}$, the effective potential $V_\text{eff}(\tilde{r})$ tends to $\operatorname{sgn} \left[(3\tilde{\gamma} - 2) \tilde{L}^2 \right] \infty$, which equals $-\infty$, since $\tilde{\gamma} \ll 1$. As $\tilde{r}$ increases toward $+\infty$, the function $V_\text{eff}(\tilde{r})$ diverges to $- \operatorname{sgn}(\tilde{k}) \infty$, which is equal to $+\infty$, since $\tilde{k} < 0$.   
	
	\subsection{Circular Orbits}
	
	Here we consider the circular motion of test particles in Weyl geometry, described by the metric~\eqref{metr} and limit ourselves to the case of those situated on the equatorial plane. 
	
	For timelike circular geodesics, the following conditions hold:
	\begin{equation}\label{co}
	    \frac{d \tilde{r}}{d \tilde{\tau}} = 0, 
	    \qquad 
	    V_{\text{eff}}(\tilde{r}_{c}) = E^{2},
	    \qquad 
	    \frac{dV_{\text{eff}}(\tilde{r})}{d \tilde{r}}\bigg|_{\tilde{r}=\tilde{r}_{c}} = 0,
	\end{equation}
	where $\tilde{\tau}=\tau/\beta$ is the dimensionless affine parameter on geodesics and  $\tilde{r}_{c} = r_{c}/\beta$ the dimensionless radius of the circular orbit, respectively.
	
	The derivative of the effective potential reads:
	\begin{equation}\label{Vp}
	    \frac{dV_{\text{eff}}(\tilde{r})}{d \tilde{r}} =
	    \frac{-2\tilde{k} \tilde{r}^{5} + \tilde{\gamma} \tilde{r}^{4} + (2-3\tilde{\gamma}-\tilde{\gamma}\tilde{L}^{2})\tilde{r}^{2} - 2(1 - 3\tilde{\gamma})\tilde{L}^{2}\tilde{r} + 3(2-3\tilde{\gamma})\tilde{L}^{2}}{2\tilde{r}^{4}}\,\,.
	\end{equation}
	
	{We}
	 note that as $\tilde{r} \rightarrow 0^{+}$, the first derivative of effective potential $dV_\text{eff}(\tilde{r})/d \tilde{r}$ tends to $\operatorname{sgn} \left[(3\tilde{\gamma} - 2) \tilde{L}^2 \right] \infty$, which equals $-\infty$, since $\tilde{\gamma} \ll 1$. As $\tilde{r}$ increases toward $+\infty$, the function $d V_\text{eff}(\tilde{r})/d \tilde{r}$ approaches $- \operatorname{sgn}(\tilde{k}) \infty$, which is equal to $+\infty$, since $\tilde{k} < 0$.   
	
	The circular orbits are the positive roots of the first derivative of the effective potential, which diverges at infinity as $\tilde{r}$ approaches $0^+$ or $+\infty$. To analyse the number of positive roots of  $dV_{\text{eff}}(\tilde{r})/d \tilde{r}=0$ we use Descartes rule for the polynomial from its numerator. In the sequence of the coefficients of the numerator, there are at most two sign changes; thus, there will be at most two circular orbits.  
	
	To find the regions in the space of the radial coordinate where the circular orbits are possible, we solve the third condition for the circular motion~(\ref{co}) with respect to the angular momentum of the test particle. We obtain:
	\begin{equation}\label{am}
	\tilde{L}= \sqrt{\frac{ -2 \tilde{k} \tilde{r}_c^3+ \tilde{\gamma} \tilde{r}_c^2 + 2 - 3 \tilde{\gamma}}{( \tilde{r}_c - 3)(\tilde{\gamma} \tilde{r}_c + 2 - 3 \tilde{\gamma})}} \cdot \tilde{r}_c\,. 
	\end{equation}
	
	{Next}
	 we analyze the relation~(\ref{am}) to find the values of $\tilde{r}_c$ where the solution for the angular momentum is well defined. For positive values of $\tilde{r}_c$, the sign of the expression under the radical is given by the first factor in the denominator, since the remaining factors are positive, being sums of terms greater than zero. Therefore, the region in which a circular motion occurs, that is, where $\tilde{L}$ defined by~(\ref{am}) exists, is given by $\tilde{r}_c>3$ or $r_c > 3 \beta$, independent of $k$ and $\gamma$. Let us mention that Turner and Horne~\cite{TH20} obtained that $r_{\text{ph}}=3 \beta$, $\beta>0$, is the unstable circular orbit of a photon. The condition for the existence of angular momentum defined by~(\ref{am}), $r_c > 3 \beta$ implies that the timelike circular geodesics orbits are outside the null circular geodesics located at $r_{\text{ph}}$.      
	We also mention that if $r_c > 3 \beta$, the corresponding energy on the circular orbit $E^2$, defined by the second condition~(\ref{co}), is well defined, as a product of two positive factors. As $r_c \rightarrow 3 \beta$, the angular momentum and the energy diverge to infinity, meaning that the unstable light sphere from $r_{\text{ph}}=3 \beta$ is the boundary of the circular orbits of the test particles.

	Next, we want to determine the radius of the last circular stable orbit for which the second derivative of the effective potential is zero. After some algebra, we get: 
	\begin{equation}\label{Vpp}
	    \frac{dV_{\text{eff}}^2(\tilde{r})}{d \tilde{r}^2} =
	    \frac{-\tilde{k}\tilde{r}^{5} + (\tilde{\gamma}\tilde{L}^{2} + 3 \tilde{\gamma} -2)\tilde{r}^{2} + 3(1-3\tilde{\gamma})\tilde{L}^{2}\tilde{r} - 6(2 - 3 \tilde{\gamma})L^{2}}{\tilde{r}^{5}}\,
	\end{equation}
	and note that as $\tilde{r} \rightarrow 0^{+}$ it tends to $-\operatorname{sgn} \left[(2-3\tilde{\gamma}) \tilde{L}^2 \right] \infty$, which equals $-\infty$, since $\tilde{\gamma} \ll 1$. As $\tilde{r}$ increases toward $+\infty$, $d V_\text{eff}^2(\tilde{r})/d \tilde{r}^2$ tend to $- \operatorname{sgn}(\tilde{k})$, which is positive, since $\tilde{k} < 0$. 
	
	Using Descartes rule for the numerator of~(\ref{Vpp}), we obtain the possible number of positive roots of  $dV_{\text{eff}}^2(\tilde{r})/d \tilde{r}^2=0$. In the sequence of its coefficients are at most three sign changes, thus there could be three or one positive solution for $dV_{\text{eff}}^2(\tilde{r})/d \tilde{r}^2=0$. For $dV_{\text{eff}}(\tilde{r})/d \tilde{r} = 0$ we found at most two solutions, and thus $dV_{\text{eff}}^2(\tilde{r})/d \tilde{r}^2=0$ has one positive~\mbox{solution}. 
	
	The innermost stable circular orbit (ISCO) is determined by the simultaneous solution of the equations:
	\begin{equation}
	\frac{dV_{\text{eff}}(\tilde{r})}{d \tilde{r}}=0, \qquad \frac{dV_{\text{eff}}^2(\tilde{r})}{d \tilde{r}^2}=0. 
	\end{equation}
	
	{Eliminating} here.
	 the angular momentum from these equations, we obtain the radius of the innermost stable circular orbit as the solution of the following quintic equation:
	\begingroup
	\makeatletter\def\f@size{9}\check@mathfonts
	\def\maketag@@@#1{\hbox{\m@th\normalsize\normalfont#1}}%
	\begin{equation}\label{eq5}
	(3 \tilde{k} \tilde{r} - \tilde{\gamma})\tilde{\gamma} \tilde{r}^4 - (3 \tilde{\gamma}-1)(8 \tilde{k} \tilde{r} - 3 \tilde{\gamma})\tilde{r}^3-(3 \tilde{\gamma}-2)\left[15 \tilde{k} \tilde{r}^3 - 6 \tilde{\gamma} \tilde{r}^2-(3 \tilde{\gamma}-1) \tilde{r} + 3 (3 \tilde{\gamma}-2)\right]=0.
	\end{equation}
	\endgroup
	
\begin{figure}[hbt!]
 \centering
 \includegraphics[width=0.55\textwidth]{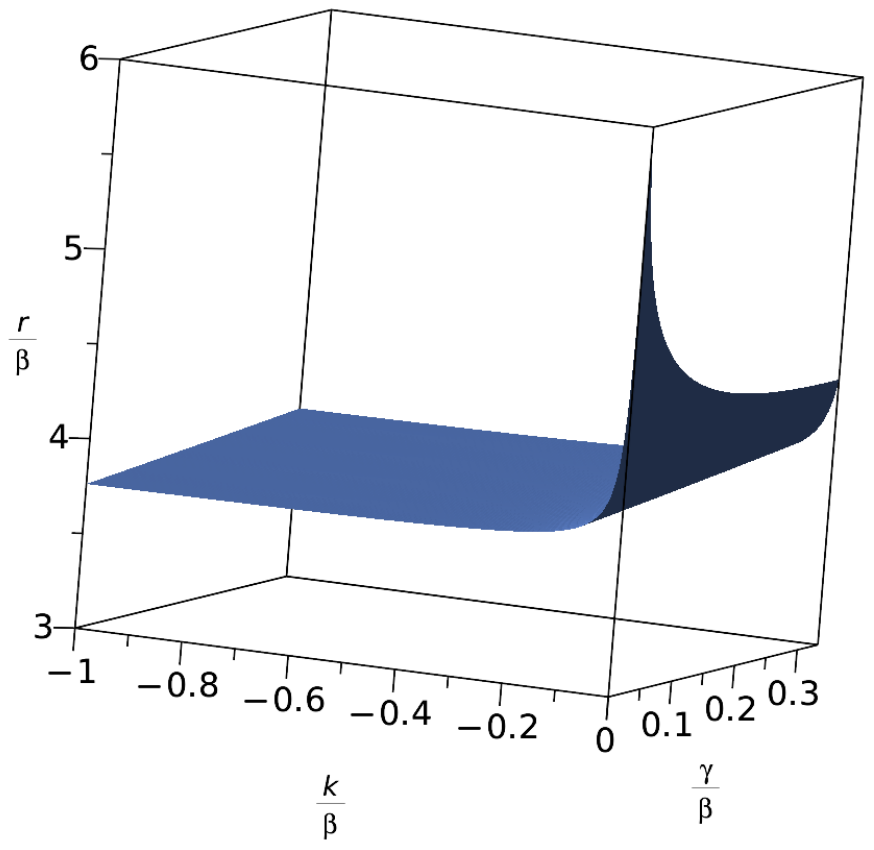}
 \caption{The innermost stable circular orbit for $\tilde{\gamma} \in [0,1/3]$ and $\tilde{k}\in [-1,0]$. We note that for $k \rightarrow 0$ and $\gamma \rightarrow 0$ the radius of the innermost circular orbit equals $r=6 \beta$, the ISCO from the Schwarzschild black hole.}\label{fig1bis}
\end{figure}

	{In}
	 the Figure~\ref{fig1bis} we present the solution of Equation~(\ref{eq5}), which give the radius of the innermost stable circular orbit for different values of $\tilde{k}$ and $\tilde{\gamma}$. 
	In the special case $\tilde{\gamma}=0$ and $\tilde{k}=0$, for which the metric~(\ref{metr}) reduces to the Schwarzschild metric, Equation~(\ref{eq5}) simplifies to $\tilde{r}-6=0$. The solution is $\tilde{r}=6$, corresponding to the dimensionless radius of the innermost stable circular orbit ($r=6 \beta$) of a Schwarzschild black hole.
	
\begin{figure}[ht!]
		\begin{center}
			\subfigure[$L=0.2$]{%
				\label{fig3:L02}
				\includegraphics[width=0.33\textwidth]{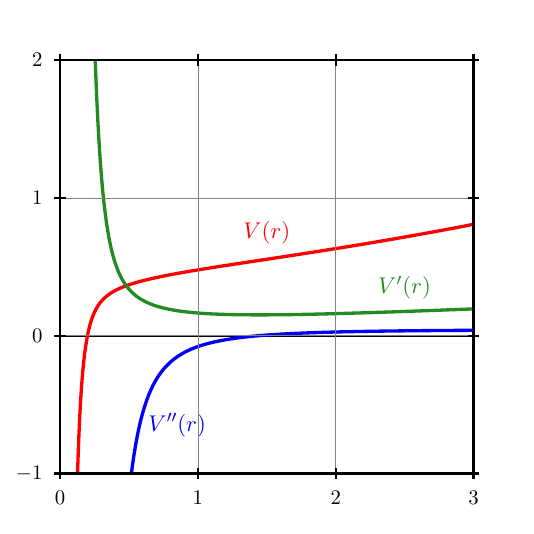}
			}%
			\subfigure[ $L=0.3793$]{%
				\label{fig3:L03793}
				\includegraphics[width=0.33\textwidth]{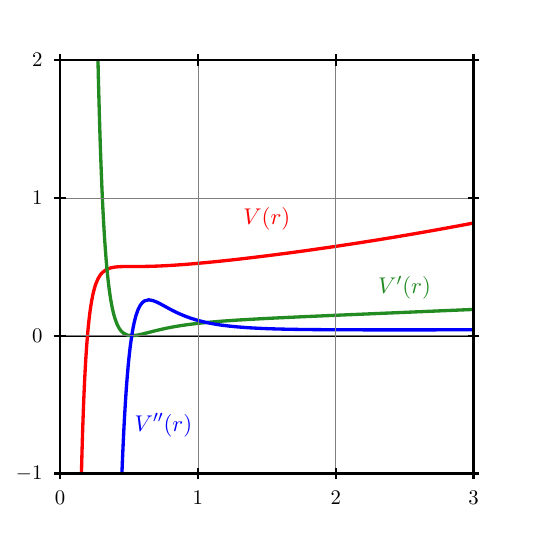}
			}%
			\subfigure[$L=0.5$]{%
				\label{fig3:L05}
				\includegraphics[width=0.33\textwidth]{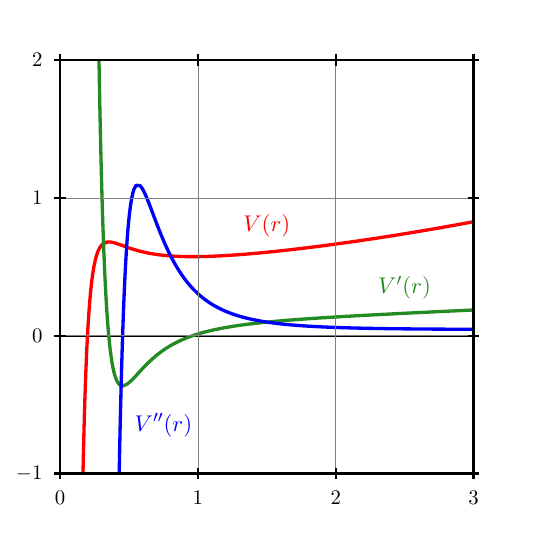}
			}%
		\end{center}
		\caption{%
			The effective potential (red line) and its derivative for $(\beta,\gamma,k) = (0.1, 0.1, -0.045)$ and three values for $L$.
			(a) If $L=0.2$, then $V'(r)>0$ for $r>0$ and no circular orbits exists.  
			(b) If $L=0.3793$ the circular orbits occur at $r = 0.52$.  
			(c) If $L=0.5$, two circular orbits occur at $r = 0.353$ and $r = 0.962$.
		}%
		\label{fig3}
	\end{figure}

	Next, we analyze the effective potential~(\ref{Ve}) and its derivatives for a fixed set of the parameters, $\beta = 0.1$, $\gamma = 0.1$, $k = -0.045$. The inflection point of the effective potential~(\ref{Ve}) is found at $L = 0.3793$. 
	In Figure~\ref{fig3}, for these values of the parameters, we plot the effective potential together with its first and second derivatives as a function of the radial coordinate $r$. An analysis of these plots shows that for $L = 0.2$, no circular orbit exists, since the first derivative of the effective potential does not vanish (see Figure~\ref{fig3}a). For $L = 0.3793$, a single circular orbit is present, corresponding to the innermost stable circular orbit, as the first derivative of the effective potential exhibits a positive double root (see Figure~\ref{fig3}b). Finally, for $L = 0.5$, two circular orbits are found, since the first derivative of the effective potential has two distinct positive roots (see Figure~\ref{fig3}c). The smaller radius corresponds to the unstable circular orbit, associated with a local maximum of the potential, while the larger radius corresponds to the stable circular orbit, where the effective potential attains a local~minimum.
	
	\section{Dynamical System Approach}\label{sect4}
	
	Differentiating Equation~\eqref{et} with respect to $\tau$ and cancelling $\dot{r}$ from both sides,\linebreak   we~obtain:
	\begin{equation}\label{2de}
	    \ddot{r} = -\frac{dV_{\text{eff}}(r)}{dr},
	\end{equation}
	a second-order differential equation that forms the starting point of the dynamical systems~analysis.
	
	\subsection{Linear Stability Analysis}
	
	The corresponding first-order system in the $(r, p)$ phase space, associated with\linebreak   Equation~\eqref{2de}, is:
	\begin{equation}\label{La}
	    \begin{cases}
		        \dot{r} = p,\\
		        \dot{p} = -V_{\text{eff}}'(r).
		    \end{cases}
	\end{equation}
	
	The Jacobian matrix of the system~\eqref{La} is:
	\begin{equation}
	    J = 
	    \begin{pmatrix}
		        0 & 1 \\
		        -V_{\text{eff}}''(r) & 0
		    \end{pmatrix},
	\end{equation}
	whose eigenvalues are:
	\begin{equation}\label{l}
	    \lambda = \pm \sqrt{-V_{\text{eff}}''(r)}.
	\end{equation}
	
	For equilibrium (fixed) points $(r_0, 0)$, Equation~\eqref{l} shows that the nature of the point depends on the sign of $V_{\text{eff}}''(r_0)$:
	
	\begin{itemize}
	    \item $(r_0, 0)$ is a \emph{{saddle point}} if $V_{\text{eff}}''(r_0) < 0$;
	    \item $(r_0, 0)$ is a \emph{{center}} if $V_{\text{eff}}''(r_0) > 0$.
	\end{itemize}
	
\begin{figure}[ht!]
		\begin{center}
			\subfigure[$L=0.2$]{%
				\label{fig4:L02}
				\includegraphics[width=0.33\textwidth]{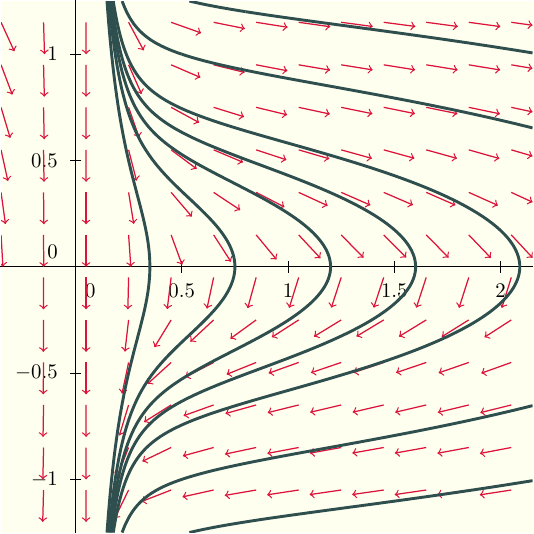}
			}%
			\subfigure[ $L=0.3793$]{%
				\label{fig4:L03793}
				\includegraphics[width=0.33\textwidth]{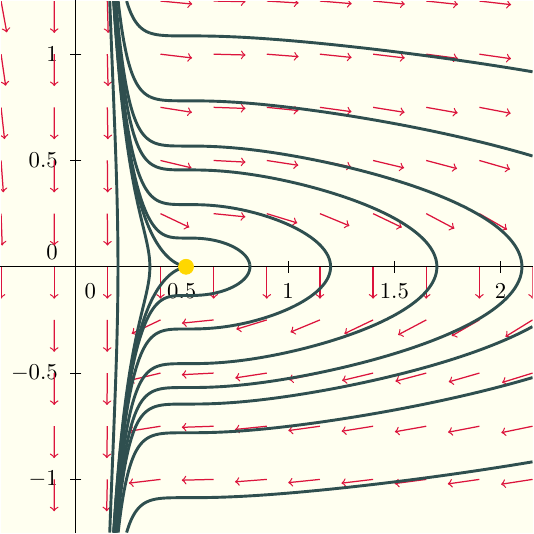}
			}%
			\subfigure[$L=0.5$]{%
				\label{fig4:L05}
				\includegraphics[width=0.33\textwidth]{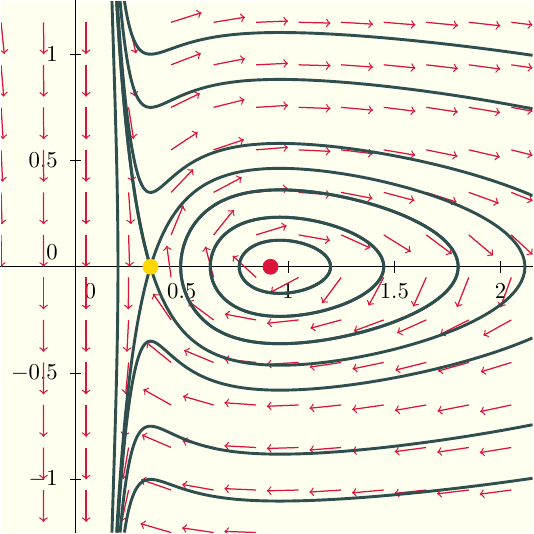}
			}%
		\end{center}
		\caption{%
			Phase portrait for $(\beta, \gamma, k) = (0.1, 0.1, -0.045)$
			and three different values for $L$.
			(a) If $L=0.2$ no equilibrium points exists.  
			(b) If $L=0.3793$, the point $(r_c,0) = (0.3024,0)$ is a cusp ($V'(r)= V''(r_{c}) = 0$).  
			(c) If $L=0.5$, the point $(r_{u},0) = (0.353,0)$ is a saddle point (gold) and $(r_{s},0) = (0.962, 0)$ is a center (red).
		}%
		\label{fig4}
	\end{figure}

	The phase portrait for the parameter set $(\beta, \gamma, k) = (0.1, 0.1, -0.045)$ and three different values of $L$ is shown in Figure~\ref{fig4}.  
	For $L = 0.2$, no circular orbit occurs (see Figure~\ref{fig4}a).  
	For $L = 0.3793$, a single circular orbit appears at $r_{c} = 0.52$ (see Figure~\ref{fig4}b).  
	The point $(r_{0}, 0) = (0.52, 0)$ is a cusp, since $V_{\text{eff}}'(r_{0}) = V_{\text{eff}}''(r_{0}) = 0$.  
	For $L = 0.5$, two circular orbits are obtained (see Figure~\ref{fig4}c):  
	$r_{u} = 0.353$ with $V_{\text{eff}}''(r_{u}) = -10.906 < 0$ and $r_{s} = 0.962$ with $V_{\text{eff}}''(r_{s}) = 0.343 > 0$. We conclude that the orbit $r_s = 0.962$ is Lyapunov stable, since $V_{\text{eff}}''(0.962) > 0$ and $r_u=0.353$ is Lyapunov unstable, since $V_{\text{eff}}''(0.353) < 0$.\vspace{-6pt}
	\subsection{Jacobi Stability Analysis}
	
	The vector field corresponding to the system~\eqref{La} is:
	\begin{equation}
	f : (0, \infty) \times \mathbb{R} \to \mathbb{R}^2, 
	\qquad f(r, p) = (p, -V_{\text{eff}}'(r)).
	\end{equation}
	
	{The}
	 stability discriminant $\Delta$ associated with the Jacobian is defined as follows:
	\begin{equation}
	    \Delta := \mathrm{tr}(J)^2 - 4\det(J) = 
	    \begin{cases}
		        < 0, & \text{Jacobi stable},\\
		        \ge 0, & \text{Jacobi unstable}.
		    \end{cases}
	\end{equation}
	{(In}
	 this particular case, the deviation curvature tensor is actually a scalar and the only eigenvalue is itself; therefore, the condition from Theorem 3 reduces to $P_1^1<0$, which is equivalent to $\Delta <0$).
	Since:
	\begin{equation}
	    \Delta\big|_{r = r_0} = 0 - 4V_{\text{eff}}''(r_0) = -4V_{\text{eff}}''(r_0),
	\end{equation}
	it follows that:
	\begin{equation}
	    \begin{cases}
		        P(r_0,0) \,\, \text{is Jacobi stable if}\ V_{\text{eff}}''(r_0) > 0,\\
		        P(r_0,0) \,\, \text{is Jacobi unstable if}\ V_{\text{eff}}''(r_0) < 0.
		    \end{cases}
	\end{equation}
	
	\section{Concluding Remarks}\label{sect5}
	
	Two complementary methods of stability analysis have been revisited here: Lyapunov (linear, in fact) stability and Jacobi stability.  
	The Lyapunov stability analysis is based on the linearization of the dynamical system near fixed points, while the Jacobi stability characterizes the behavior of nearby trajectories as a whole.  
	 The KCC theory also allows a direct geometrization of dynamical systems, which thus can be described in purely geometric terms through the introduction of a non-linear connection, of an associated covariant derivative, and the curvature deviation tensor, respectively.
	
	We have shown that, for circular orbits in Weyl conformal gravity, the condition for Jacobi stability coincides with that for linear Lyapunov stability.  
	Thus, in this context, the two notions of stability are equivalent.
	
	The equivalence between Lyapunov and Jacobi stability for circular timelike geodesics around spherically symmetric Weyl black holes provides an important physical insight.  
	Both analyses indicate that the local curvature governs not only the linear response at small perturbations but also the behaviour of nearby trajectories. 
	
	To conclude, in the present paper, we have introduced and studied in detail some geometrical methods necessary for the analysis and description of the Jacobi and Lyapunov stability properties of black hole solutions obtained in the framework of the conformally invariant Weyl gravity. The approaches considered in the present analysis may be extended easily to other classes of black hole solutions obtained in modified gravity theories. Such methods may lead to a better understanding of the dynamical properties and stability of the physical processes taking place around black holes.
	
\bigskip

\textbf{Acknowledgments:} The authors thank Tiberiu Harko for discussions that helped us to improve our work. The authors also want to thank the referees who helped us to improve the quality and clarity of our work. This paper is an extension of the contribution we presented at the XIIIth Bolyai-Gauss-Lobachevsky Conference held in Sa\"{\i}dia, Morocco, in May 26--29, 2025. C.B. and P.B. acknowledge support from the 2024 Development Fund of Babes-Bolyai University for funding the conference participation.


\begin{thebibliography}{999}

\bibitem{Will} Will, C.M. The Confrontation between General Relativity and Experiment. {\em Living Rev. Relativ.} {\bf 2014}, {\em 17}, 4. 

\bibitem{Mar} De Marchi, F.; Cascioli, G. Testing general relativity in the solar system: present and future perspectives. {\em Class. Quantum Gravity} {\bf 2020}, {\em 37}, 095007.

\bibitem{grad1} Abbott, B.; Abbott, R.; Abbott, T.; Abernathy, M.; Acernese, F.; Ackley, K.; Adams, C.; Adams, T.; Addesso, P.; LIGO Scientific Collaboration and Virgo Collaboration. Observation of Gravitational Waves from a Binary Black Hole Merger. {\em Phys. Rev. Lett.} {\bf 2016}, {\em 116}, 061102. 

\bibitem{grad2} Abbott, B.; Abbott, R.; Abbott, T.; Abernathy, M.; Acernese, F.; Ackley, K.; Adams, C.; Adams, T.; Addesso, P.; LIGO Scientific Collaboration and Virgo Collaboration. Properties of the Binary Black Hole Merger GW150914.  {\em Phys. Rev. Lett.} {\bf 2016}, {\em 116}, 241102.

\bibitem{acc1} Riess, A.G.; Filippenko, A.V.; Challis, P.; Clocchiatti, A.; Diercks, A.; Garnavich, P.M.; Gilliland, R.L.; Hogan, C.J.; Jha, S.; Kirshner, R.P.; et al. Observational Evidence from Supernovae for an Accelerating Universe and a Cosmological Constant. {\em Astron. J.} {\bf 1998}, {\em 116}, 1009--1038. 

\bibitem{acc2}  Perlmutter, S.; Aldering, G.; Goldhaber, G.; Knop, R.A.; Nugent, P.; Castro, P.G.; Deustua, S.; Fabbro, S.; Goobar, A.; Groom, D.E.; et al. Measurements of $\Omega$ and $\Lambda$ from 42 High-Redshift Supernovae. {\em Astrophys. J.} {\bf 1999},  {\em 517}, 565--586. 

\bibitem{Amen} Amendola, L.; Tsujikawa, S. {\em  Dark Energy}; Cambridge University Press: Cambridge, UK, {2010}.

\bibitem{DM1} Oks, E. Review of latest advances on dark matter from the viewpoint of the Occam razor principle. {\em New Astron. Rev.} {\bf 2023}, {\em 96},~101673. 

\bibitem{DM2}  Bian, L.; Liu, X.; Xie, K.-P. Probing superheavy dark matter with gravitational waves. {\em J. High Energy Phys.} {\bf 2021}, {\em 2021}, 175.

\bibitem{HTens} Valentino, E.D.; Mena, O.; Pan, S.; Visinelli, L.; Yang, W.; Melchiorri, A.; Mota, D.F.; Riess, A.G.; Silk, J. In the realm of the Hubble tension—A review of solutions. {\em  Class.  Quantum Gravity} {\bf 2021}, {\em 38}, 153001.       

\bibitem{W18a} Weyl, H. {Gravitation and electricity}.
{\em Sitzungsber. Preuss. Akad. Wiss. Berlin (Math. Phys.)} {\bf 1918}, Erster Halbband (Januar bis Juni)  {465--480.}


\bibitem{W18b} Weyl, H. Reine Infinitesimalgeometrie. {\em Math. Z}. {\bf 1918}, {\em 2}, 384--411.

\bibitem{Scholtz} Scholz, E. The unexpected resurgence of Weyl geometry in late 20-th century physics. \emph{arXiv} \textbf{2017}, arXiv:1703.03187.

\bibitem{B21} Bach, R. Zur Weylschen Relativit\"atstheorie und der Weylschen Erweiterung des Kr\"ummungstensorbegriffs.
{\em Math. Z.} {\bf 1921}, {\em 9},~110--135.

\bibitem{MK89} Mannheim, P.D.; Kazanas, D. Exact vacuum solution to conformal Weyl gravity and galactic rotation curves. {\em Astrophys. J.} {\bf 1989},{\em 342},~635--638.

\bibitem{KM91} Kazanas, P.D.; Mannheim, P.D. 
General Structure of the Gravitational Equations of Motion in Conformal Weyl Gravity. {\em Astrophys. J. Suppl.  Ser.} {\bf 1991}, {\em 76}, 431--453.

\bibitem{M92} Mannheim, P.D. Conformal Gravity and the Flatness Problem. {\em Astrophys. J.} {\bf 1992}, {\em 391}, 429--432. 

\bibitem{M96} Mannheim, P.D. Conformal cosmology and the age of the universe. \emph{arXiv}  {\bf 1996}, 	arXiv:astro-ph/9601071. 

\bibitem{Dirac73}Dirac, P.A.M. Long range forces and broken symmetries. \emph{Proc. R. Soc. Lond. A} \textbf{1973}, \emph{333}, 403--418.

\bibitem{Ghil19} Ghilencea, D.M.; Lee, H.M. Weyl gauge symmetry and its spontaneous breaking
in the standard model and inflation. \emph{Phys. Rev. D} \textbf{2019}, \emph{99}, 115007. 

\bibitem{lessa25} Lessa, L.A.; Macedo, C.F.B.; Ferreira, M.M., Jr. Black holes in higher-derivative Weyl conformal gravity. \emph{arXiv} \textbf{2025}, arXiv:2509.21495v1.

\bibitem{EP98} Edery, A.;  Paranjape, M.B. Classical tests for Weyl gravity: Deflection of light and time delay. {\em Phys. Rev. D} {\bf 1998},  {\em 44}, 417--423. 

\bibitem{SKS12} Sultana, J.; Kazanas, D.; Said, J.L. Conformal Weyl gravity and perihelion preccesion. {\em Phys. Rev. D} {\bf 2012}, {\em 86}, 084008.

\bibitem{TH20} Turner, G.E.; Horne, K. Null geodesics in conformal gravity. {\em Class. Quantum Gravity} {\bf 2020}, {\em 37}, 095012.

\bibitem{Ko33} Kosambi, D.D. Parallelism and path-spaces. {\em Math. Z.} {\bf 1933}, {\em 608}, 608--618.

\bibitem{Ca33} Cartan, E. Observations sur le m\'{e}moir pr\'{e}c\'{e}dent. {\em Math. Z.} {\bf 1933}, {\em 37}, 619--622.

\bibitem{Ch39} Chern, S.S. Sur la g\'{e}om\'{e}trie d’un syst\'{e}me d’equations differentialles du second ordre. {\em Bull.  Sci. Math.} {\bf 1939}, {\em 63}, 206--212.

\bibitem{rev} Boehmer, C.G.; Harko, T.; Sabau, S.V. Jacobi stability analysis of dynamical systems---Applications in gravitation and cosmology. {\em Adv. Theor. Math. Phys.} {\bf 2012}, {\em 16}, 1145--1196.

\bibitem{X2} Abolghasem, H. Jacobi Stability of Circular Orbits in a Central Force. {\em J. Dyn. Syst. Geom. Theor.} {\bf 2012}, {\em 10}, 197--214.

\bibitem{X3} Abolghasem, H.  Stability of circular orbits in Schwarzschild spacetime. {\em Int. J. Differ. Equ. Appl.} {\bf 2013}, {\em 12}, 131--147.

\bibitem{BBH2} Blaga, P.; Blaga, C.; Harko, T. Jacobi and Lyapunov Stability Analysis of Circular Geodesics around a Spherically Symmetric Dilaton Black Hole. {\em Symmetry} {\bf 2023}, {\em 15}, 329. 

\bibitem{Ha6} Lake, M.J.; Harko, T. Dynamical behavior and Jacobi stability analysis of wound strings. {\em Eur. Phys. J. C} {\bf 2016}, {\em 76}, 311.

\bibitem{BBH1} Blaga, C.; Blaga, P.; Harko, T. Jacobi stability analysis of the classical restricted three body problem. {\em Rom. Astron. J.} {\bf 2021}, {\em 31},~101--112. 

\bibitem{An03} Antonelli, P.L.; Buc\u{a}taru, I. KCC theory of system of second order differential equations. In  \textit{Handbook of Finsler Geometry};\linebreak   Antonelli, P.L., Ed.; Kluwer Academic: Dordrecht, The Netherlands, {2003}; Volume 1, pp.~83--174.

\bibitem{LC27}Levi-Civit\'a, T. Sur l'\'ecart g\'eod\'esique. \emph{Math. Ann.} \textbf{1927}, \emph{97}, 291--320.

\bibitem{Syn27}Synge, J.L. On the geometry of dynamics. \emph{Philos. Trans. R. Soc. Lond. A} \textbf{1927}, \emph{226}, 31--106.

\bibitem{AnIM93} Antonelli, P.L.; Ingarden, R.S.; Matsumoto, M. \textit{The Theory of Sprays and Finsler Spaces with Applications in Physics and Biology}; Springer Science+Business Media: Dordrecht, The Netherlands, {1993}; Volume~58.

\bibitem{An01} Antonelli, P.L.; Bucataru, I. New results about the geometric invariants in KCC theory. \emph{An. Stiint. Univ. Al. I. Cuza Iasi. Mat. (N.S.)} {\bf 2001}, {\em 47}, 405--420.

\bibitem{H67} Hahn, W. {\em Stability of Motion}; Springer: Berlin, Germany, {1967}.

\bibitem{HSD} Hirsch, M.W.; Smale, S.; Devaney, R.L.  \emph{Differential Equations, Dynamical Systems and an Introduction to Chaos}, 3rd ed.;  {Academic Press, Waltham, MA,USA} 
{2013}.
\bibitem{M06}Mannheim, P.D.  Alternatives to Dark Matter and Dark Energy. \emph{Prog. Part. Nucl. Phys.} \textbf{2006}, \emph{56}, 340--445.

\bibitem{nak2003}Nakahara, M.  \emph{Geometry, Topology and Physics}; IOP Publishing: {Bristol, UK,}
{2003}; p. 462. 

\bibitem{Lan38}Lanczos, C.  A Remarkable Property of the Riemann-Christoffel Tensor in Four Dimensions. \emph{Ann.  Math.} \textbf{1938}, \emph{39}, 842--850.

\bibitem{cond25}Condeescu, C.; Micu, A.  The gauge theory of Weyl group and its interpretation as Weyl quadratic gravity. \emph{ Class. Quantum Grav.} \textbf{2025}, \emph{42}, 065011.


\bibitem{MO13} Mannheim, P.D.; O'Brien, J.G. Galactic rotation curves in conformal gravity. {\em J. Phys. Conf. Ser.} {\bf 2013}, {\em 437}, 012002. 

\bibitem{LD23} Lungu, V.; Dariescu, M.-A. Circular geodesics in Manheim-Kazanas spacetime. {\em Proc.  Rom. Acad. Ser. A} {\bf 2023}, {\em 24}, 35--42. 




\end{thebibliography}
\end{document}